\documentclass[fleqn,10pt]{wlscirep}

\title{Evolution of the electronic band structure of twisted bilayer graphene upon doping}

\author[1]{Shengqiang Huang}
\author[1]{Matthew Yankowitz}
\author[1]{Kanokporn Chattrakun}
\author[1,2]{Arvinder Sandhu}
\author[1,*]{Brian J. LeRoy}
\affil[1]{Physics Department, University of Arizona, Tucson, AZ 85721, USA}
\affil[2]{College of Optical Sciences, University of Arizona, Tucson, AZ 85721, USA}

\affil[*]{leroy@physics.arizona.edu}

\begin{abstract}
The electronic band structure of twisted bilayer graphene develops van Hove singularities whose energy depends on the twist angle between the two layers.  Using Raman spectroscopy, we monitor the evolution of the electronic band structure upon doping using the G peak area which is enhanced when the laser photon energy is resonant with the energy separation of the van Hove singularities. Upon  charge doping, the Raman G peak area initially increases for twist angles larger than a critical angle and decreases for smaller angles.  To explain this behavior with twist angle, the energy of separation of the van Hove singularities must decrease with increasing charge density demonstrating the ability to modify the electronic and optical properties of twisted bilayer graphene with doping.
\end{abstract}
\begin{document}

\flushbottom
\maketitle
\thispagestyle{empty}

%\noindent Please note: Abbreviations should be introduced at the first mention in the main text – no abbreviations lists. Suggested structure of main text (not enforced) is provided below.

\section*{Introduction}

Twisted bilayer graphene (tBLG) has drawn a lot of theoretical attention because of its unique electronic band structure, which depends on the twist angle between the two graphene monolayers\cite{Santos2007,Magaud2010,Shallcross2010,Bistritzer2011}.  The twist angle between the layers causes the Dirac cones of the individual layers to be displaced in momentum. When the linear bands from the two monolayer graphene sheets cross, saddle points appear in the band structure of tBLG, giving rise to logarithmic van Hove singularities (vHs) in the density of states (DoS)\cite{VanHove1953}. Twisted bilayer graphene occurs in many different graphene preparation methods such as chemical vapor deposition (CVD) growth on metals\cite{Li2009,Lu2013,Nie2011,Murata2012,Iwasaki2014}, precipitation from silicon carbide\cite{Varchon2008, Hass2008} and mechanical folding of monolayer graphene\cite{Carozo2013}. Scanning tunneling microscopy (STM) and spectroscopy have been employed to explore the properties of tBLG, where a periodic potential leads to a moir\'e pattern in the topography and vHs in the local DoS\cite{LiGuohong2010, Luican2011, Brihuega2012, He2012}. Raman spectroscopy is another important characterization tool that is faster than STM and nondestructive to the sample\cite{Ferrari2006}. The Raman spectrum of carbon based materials contains two main peaks, namely the G peak around 1580 cm$^{-1}$ and the 2D peak around 2700 cm$^{-1}$ \cite{Ferrari2006,Malarda2009}. It has been observed that the position, intensity, and width of both the Raman G and 2D peaks depend on the twist angle\cite{Poncharal2008,Ni2008,ZNi2009,Kim2012, Havener2012}. In addition, new peaks appear in the Raman spectrum of tBLG due to extra scattering pathways enabled by the periodic potential\cite{Carozo2013,Gupta2010,Carozo2011,JCampos2013,He2013,Jorio2013}. Theoretical calculations show that variations in the Raman peaks with twist angle are due to the angle dependent electronic and phononic band structures.\cite{Sato2012,Cocemasov2013,Coh2013}.

The band structure of tBLG is predicted to depend on the potential difference between the two layers\cite{Santos2007,Xian2011,Jose2013,Moon2014}, which can be generated by charge doping or electrostatic gating.  Recently in tBLG, G peak splitting and quenching upon doping have been observed for some specific twist angles\cite{Chiu2014,Chung2015}.  However, it is not clear if this is a universal behavior for all twist angles.  Therefore, we have conducted a systematic Raman study of tBLG for a variety of twist angles, using a 532 nm laser excitation, with controllable charge densities up to $10^{13}$ cm$^{-2}$. The variation of the intensity of the Raman G peak of tBLG as a function of charge density demonstrates that G peak quenching is not a universal behavior but rather a consequence of the electronic band structure depending on density.  This gives rise to different variations of the G peak intensity with charge density depending on twist angle. 

\section*{Experimental setup} 
\begin{figure}[h]
\begin{center}
\includegraphics[width=12cm]{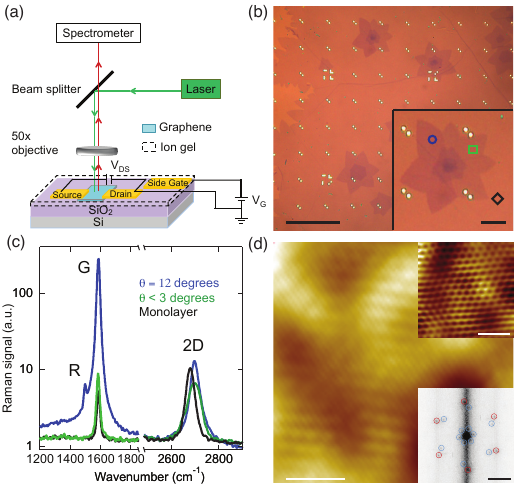} 
\caption{(a) Schematic diagram of the experimental layout. (b) Optical microscope image of CVD grown graphene flakes. Bright dots are gold coordinate markers on the SiO$_2$/Si substrate. The inset is a zoomed in image of a multilayer graphene flake. Blue circle, green square and black diamond are locations where the Raman spectra in (c) are acquired.  Scale bars in the main image and the inset are 50 $\mu$m and 10 $\mu$m respectively. (c) Raman spectra at the marked locations in (b). The black diamond is on monolayer graphene while the other two locations are from tBLG. Only wavenumber ranges near the G and 2D peaks are shown.(d) STM topography image of tBLG showing a 1.19 $\pm$ 0.03 nm moir\'e pattern. Sample voltage is 0.3 V and tunnel current is 100 pA.  Scale bar in the figure is 4 nm.  The upper inset is a zoom-in image showing the atomic lattice. Scale bar is 1 nm. The lower inset is the FFT of the topography showing the atomic lattice and moir\'e pattern features, red and blue circles respectively. Scale bar is 10 nm$^{-1}$. }
\label{fig:RamanOnCVD}
\end{center}
\end{figure}

%\subsection*{Subsection}
The experimental layout for the charge doping measurements is sketched in Fig.~\ref{fig:RamanOnCVD}(a). Raman spectroscopy was performed on a home built Raman system using a 532 nm Nd:YAG laser under ambient environment and room temperature (see details in
the Method section). For selected tBLG flakes, source, drain and side gate electrodes were written by electron beam lithography followed by deposition of Cr/Au (5 nm/30 nm). A polymer electrolyte top gate was employed to induce charge carriers in the sample \cite{LiuJie2004}(see details in the Method section). Debye layers are formed near the graphene and the side gate electrode where ions accumulate. The thickness of these Debye layers are typically a few nanometers\cite{LiuJie2004,Das2008} so that the capacitance is significantly larger than that of the Si back gate. Side gate voltages applied to the device were kept to between -1.5 V and 3.0 V to avoid any electrochemical reactions in the polymer electrolyte. The gate voltage was constant during the acquisition of the Raman spectra and then was changed slowly at a rate of 1 mV/s between measurements to ensure the polymer electrolyte remained stable. The leakage current was monitored during the measurements and there was no detectable signal.

The graphene in this study was grown via a low pressure CVD method on  25 $\mu$m thick Cu foil placed within a sealed copper pouch\cite{Li2011}(see details in the Method section). Our growth conditions yield large monolayer graphene flakes with smaller multilayer regions originating from the nucleation sites. Fig.~\ref{fig:RamanOnCVD}(b) shows an optical microscope image of the graphene flakes after they have been transferred from the Cu foil onto an oxidized Si substrate (285 nm SiO$_2$ thickness) using a wet transfer method\cite{LiTra2009}. The number of layers is clearly distinguishable based on the optical contrast. The size of the monolayer region is over 200 $\mu$m on a side, while the bilayer is on the order of 25 $\mu$m and the multilayer region in the center is smaller than 10 $\mu$m. The contrast varies on different regions of the bilayer, which indicates the angle between the first and second layer varies due to different domains in the CVD grown graphene.  Specifically, the slightly darker areas (e.g. blue circle in the inset of Fig.~\ref{fig:RamanOnCVD}(b)) correspond to approximately a 12 degree twist angle which absorbs more visible light owing to the enhanced DoS in this energy range\cite{Moon2013,Tabert2013}.
%%%%%%%%%%%%%%%
%%%%%%%%%%%%%%%%%%

\section*{Twist angle determination by STM and Raman measurements}

While previous studies have employed contrast measurements to identify the twist angle\cite{Robinson2013,Campos2013}, we find Raman spectroscopy to be a more precise and convenient method.  Raman spectra taken at the locations marked in the inset of Fig.~\ref{fig:RamanOnCVD}(b) are plotted in Fig.~\ref{fig:RamanOnCVD}(c). There is no D peak visible in the spectra either before or after the polymer electrolyte measurements which indicates that the measured area is free of defects\cite{Ferrari2006}. In tBLG, both the Raman G and 2D peaks have been found to vary with the twist angle as measured with transmission electron microscopy\cite{Kim2012, Havener2012}.  Remarkably, on the slightly darker region (blue curve), which has a twist angle of 12 degrees, the G peak intensity is 25 times larger than on the lighter region that has a twist angle less than 3 degrees (green curve). Both bilayer spectra are reliably distinguishable from that of monolayer graphene marked by the black diamond (black curve). For the 532 nm laser in our study, the G peak is enhanced for twist angles from 10 to 15 degrees. 

The twist angle between the two layers of tBLG leads to a moir\'e pattern, whose wavelength depends inversely on the twist angle $\theta$.  Therefore, measuring the wavelength of the moir\'e pattern using STM gives a direct measurement of the twist angle. All topography measurements were performed in a low-temperature ultrahigh vacuum STM at a temperature of 4.5 K.  These measurements were performed before applying the polymer electrolyte to the device. Fig.~\ref{fig:RamanOnCVD}(d) shows a typical topography image of tBLG where a 1.19 $\pm$ 0.03 nm moir\'e pattern is visible. The upper inset of Fig.~\ref{fig:RamanOnCVD}(d) is a zoomed in topography image showing the graphene atomic lattice.  A Fourier transform of the topography image gives the wavevectors of the atomic lattice and moir\'e pattern. The lower inset of Fig.~\ref{fig:RamanOnCVD}(d) is a FFT of the topography with the atomic lattice and moir\'e pattern wavevectors marked by red and blue circles respectively. The ratio between the atomic lattice length $a$ and the moir\'e pattern wavelength $D$ gives the twist angle $\theta$ between the two lattices using $a/D = 2 \sin(\theta/2)$.

%%%%%%%%%%%%%%%
\begin{figure}[h]
\begin{center}
\includegraphics[width=12cm]{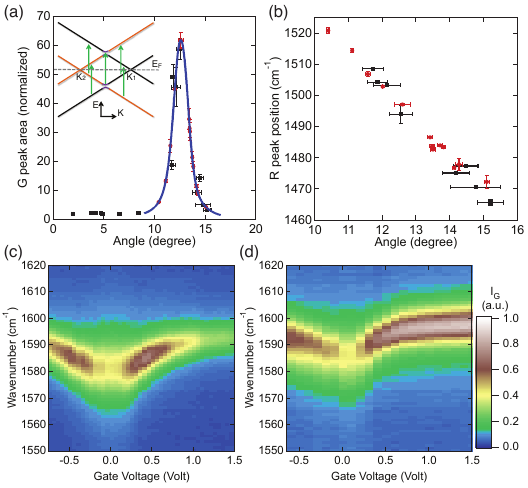} 
\caption{(a) Measured Raman G peak area (normalized to that of monolayer) as a function of twist angle. The twist angle for the black points was measured using STM topography. The uncertainty in angle of the black points represents the variation in the atomic lattice length $a$ and moir\'e pattern wavelength $D$ measured by STM.  The blue curve is the theoretical calculation within the angle range where the G peak is enhanced showing that the maximum enhancement occurs at 12.6$^{\circ}$. The additional red points were only measured by Raman spectroscopy and their twist angles were determined as discussed in the text. The error bars in G peak area represent the standard deviation of the area of a Lorentzian fit to the G peak. Inset shows Raman pathways that contribute to the enhanced G peak. (b) Raman R peak position as a function of twist angle. The error bars in R peak position represent the uncertainty of the peak position when fit with a Lorentzian. (c)-(d) Raman intensity near the G peak as a function of wavenumber and gate voltage for 11.1$^{\circ}$(c) and 14.1$^{\circ}$(d) respectively.} 
\label{fig:Raman}
\end{center}
\end{figure}
%%%%%%%%%%%%%%%%%%

Using the above procedure, more than 15 different tBLG flakes were imaged using STM to determine their twist angles.  Fig.~\ref{fig:Raman}(a) plots the measured Raman G peak area of these tBLG flakes normalized to monolayer graphene.  There is a clear enhancement of the G peak area near a twist angle of 12 degrees. Our data agrees with previous results\cite{Kim2012,Havener2012}, where the twist angle was determined by transmission electron microscopy.  The enhancement at a specific angle range arises due to the enhanced joint density of states at the photon energy of the laser excitation.

To quantify the enhancement, the Raman G peak area as a function of twist angle is calculated using\cite{Falicov1983}
%%%%%%%%%%%%%%% 
\begin{equation}
\label{RamanGpeakCalculation}
\\A \propto \left| \sum_{k} \frac{\left| M_{\mathrm{op}} \right|^2 M_{\mathrm{el-ph}}}{(\Delta E(k)-i\gamma)(\Delta E(k)-E_{\mathrm{ph}}-i\gamma)}\right| ^2
\end{equation}
%%%%%%%%%%%%%%% 
where $\Delta E(k) = E_{\mathrm{ex}}-E_{\mathrm{ab}}(k)$, $E_{\mathrm{ex}}=2.33$ eV is the photon energy of the laser excitation, $E_{\mathrm{ab}}(k)$ is the energy separation of electronic states a and b, $E_{ph} = 0.196$ eV is the G peak phonon energy, $\gamma$ is the inelastic scattering rate for which we use 0.15 eV, $M_{\mathrm{el-ph}}$ is the electron-phonon coupling matrix element, and $M_{op}$ is the optical transition matrix element. We follow reference \cite{Santos2007} to calculate the electronic band structure of tBLG, which retains the linear dispersion near the Dirac points and exhibits vHs and splitting where the bands from the two layers meet. We include only the four lowest energy bands to perform the calculation as depicted in the inset of Fig.~\ref{fig:Raman}(a).  The largest enhancement occurs when the photon energy of the laser is on resonance with the energy separation of a large number of electronic states as denoted by the green arrows in the inset of Fig.~\ref{fig:Raman}(a). $M_{\mathrm{el-ph}}$ is taken as a constant in the calculation as we are only interested in changes in the band structure due to twist angle variations. Furthermore, the twist angle corresponding to the maximum enhancement depends on laser photon energy and therefore $M_{\mathrm{el-ph}}$ can not be responsible for the observed enhancement\cite{Kim2012,Havener2012}. $M_{op}$ along the line passing through the two Dirac points is modeled by a phenomenological Lorentzian as in reference\cite{Havener2012} with no k-space dependence in the orthogonal direction.  The blue curve in Fig.~\ref{fig:Raman}(a) plots the result of the calculation which matches the experimentally measured data.

The center position of the peak in the calculated curve is at 12.6 degrees, which is defined as the critical angle.  Using a different photon energy shifts the critical angle.  The calculation allows a Raman spectroscopy measurement of the G peak area to determine two possible twist angles for the tBLG flake.  To distinguish between the two possible twist angles, we use the Raman R peak associated with the transverse optical (TO) phonon, which emerges from a double-resonance intervalley process mediated by the periodic potential\cite{Carozo2011}.  In the angle range where the G peak is enhanced, the R peak is clearly visible, as shown in Fig.~\ref{fig:RamanOnCVD}(c). The R peak position decreases linearly with increasing twist angle, -11.7$\pm$0.8 $\mathrm{cm^{-1}/degree}$, as shown in Fig.~\ref{fig:Raman}(b) which allows the twist angle to be uniquely identified.  Again, the black circles correspond to tBLG flakes whose twist angles were measured by STM. The red circles correspond to tBLG flakes whose twist angles were inferred from Raman spectroscopy measurements of the G peak area. Using a combination of the R peak position and the G peak area, the twist angle can be determined solely from Raman measurements.

\section*{Charge doping induced by a polymer top gate}

To study the charge density dependence of the Raman G peak, we use a voltage on the polymer gate, $V_g$ to induce charge carriers in the tBLG. As the charge density changes, the G peak behaves very differently depending on the twist angle as shown in Fig.~\ref{fig:Raman}(c) and ~\ref{fig:Raman}(d). These data are from two different tBLG flakes which both have approximately 12 times enhancement at zero gate voltage but two different twist angles. In both cases, the position of the G peak shifts to higher wavenumbers as seen in monolayer graphene~\cite{Pisana2007,Yan2007} and previous Raman studies of tBLG~\cite{Poncharal2008,Chung2015}. Similarly, the 2D peak position for all tBLG flakes shows the same behavior as monolayer graphene where it is initially constant then shifts to higher wavenumbers for hole doping and lower wavenumbers for electron doping\cite{Poncharal2008,Chung2015}.  While these aspects of the measurements are the same for all twist angles, the intensity of the G peak shows a strong twist angle dependence. As the gate voltage increases the G peak intensity initially increases but then quickly decreases in Fig.~\ref{fig:Raman}(c) while it continually increases in Fig.~\ref{fig:Raman}(d). To understand why these two different flakes behave so differently with gate voltage, we must examine the effect of the voltage on the band structure of the tBLG.

In monolayer graphene, the voltage on the polymer gate acts to both shift the Fermi level and produce a potential between the graphene and electrode.  The Fermi level is related to the charge density $n$ by $E_F = \hbar v_F \sqrt{\pi n}$, where $\hbar$ is the reduced Planck constant and $v_F = 1.05 \times 10^6$ m/s is the Fermi velocity of  monolayer graphene\cite{Das2008}.  The potential between the gate electrode and graphene is given by $U = \varepsilon_{PE}E/C_g = ne/C_g$, where $\varepsilon_{PE}$ is the dielectric constant of the polymer electrolyte, $E$ is the electric field and $C_g$ is the geometric capacitance per unit area of the Debye layer in the polymer electrolyte. Combining these two effects gives, 
%%%%%%%%%%%%%%% 
\begin{equation}
\label{MonolayerChargeDensity}
\\eV_g = E_F + e U  = \hbar v_F \sqrt{\pi n} + \frac{e^2}{C_g} n.
\end{equation}
%%%%%%%%%%%%%%%
Here $V_g$ is measured relative to the voltage when the Fermi level is at the Dirac point by finding at which gate voltage the G peak position is a minimum.  To determine the geometric capacitance, we measured the G peak position for monolayer graphene as a function of the applied gate voltage. The shift of the G peak $\Delta \Omega_G$ is proportional to the Fermi energy, namely, $\Delta \Omega_G = \alpha E_F$, where $\alpha$ is 42 cm$^{-1}$ eV$^{-1}$ taken from ref.~\cite{Wang2011}.  For each device, $\Delta \Omega_G$  versus gate voltage was measured to determine $C_g$.  For the different devices measured, the capacitance per unit area varied within the range $0.7 - 2.2\; \mu$F cm$^{-2}$, which is about two orders of magnitude larger than that of the Si back gate. The geometric capacitance per unit area for bilayer graphene areas was taken to be the same as that of nearby monolayers since the polymer electrolyte was uniformly distributed over these length scales. The doping at the position of the laser spot was directly measured and therefore global spatial variations do not affect the results.

In tBLG, there is an additional effect because the top layer is more heavily doped than the bottom layer due to its proximity to the polymer electrolyte~\cite{Chung2015}.  This asymmetric doping causes an interlayer potential $\Delta\phi$ to develop. A simple model is adopted from ref.~\cite{Chung2015} to calculate the charge densities $n_T$ and $n_B$ in the top and bottom layers respectively. The tBLG is modeled as a parallel plate capacitor with two weakly coupled monolayers separated by an interlayer distance $d_0 = 0.34$ nm.  This model is valid as long as the doping level is small compared to the separation of the vHs as is the case in our experiment.  The top and bottom layers experience different electric fields, given as $E_T = e (n_T+n_B)/\varepsilon_{PE}$ and $E_B = e n_B/\varepsilon_{tBLG}$, where $\varepsilon_{tBLG}$ is the dielectric constant of the tBLG.  Similar to equation~\ref{MonolayerChargeDensity} in the monolayer case, for bilayer we use the charge density and electric field of the top layer to obtain, 
%%%%%%%%%%%%%%% 
\begin{equation}
\label{BilayerChargeDensity}
\\eV_g = \hbar v_F \sqrt{\pi n_T} + \frac{e^2}{C_g} (n_T + n_B).
\end{equation}
%%%%%%%%%%%%%%%

The interlayer potential energy $e\Delta\phi$ is equal to the energy difference between the Dirac points of the top and bottom layers $e\Delta\phi = \hbar v_F \sqrt{\pi n_T}-\hbar v_F \sqrt{\pi n_B}$. In the parallel plate capacitor model, the interlayer potential is simply the electric field between the layers $E_B$ times their separation $d_0$. Namely, $\Delta\phi = e n_B/C_{tBLG}$, where $C_{tBLG} = \varepsilon_{tBLG}/d_0$ is the effective interlayer capacitance per unit area. Combining the above two equations we obtain,  
\begin{equation}
\label{DeltaPhi}
\\ \hbar v_F \sqrt{\pi n_T}-\hbar v_F \sqrt{\pi n_B}= \frac{e^2}{C_{tBLG}} n_B
.
\end{equation}
%%%%%%%%%%%%%%%
$n_T$ and $n_B$ can be solved for any value of $V_g$ using equations~\ref{BilayerChargeDensity} and ~\ref{DeltaPhi} and the measured $C_g$. We take $C_{tBLG} = 8 \times 10^ {-6}$ F cm$^{-2}$  in our calculation \cite{Elton2013}. In the inset of Fig.~\ref{fig:results}(a), we use these equations to convert the applied gate voltage, $V_g$ to the charge density $n_T$(black) in the top layer, $n_B$(red) in the bottom layer and $n_{Total}$(blue) for the tBLG corresponding to Fig.~\ref{fig:Raman}(c).  Once the conversion from gate voltage to charge density is known, the area of the G peak is fit for each gate voltage with a single Lorentzian to obtain the blue curve in Fig.~\ref{fig:results}(a).  A similar procedure is used for all other tBLG flakes to obtain the remaining curves in Figs.~\ref{fig:results}(a) and (b). 

\section*{Raman G peak area dependence on charge doping}

\begin{figure}[h]
\begin{center}
\includegraphics[width=12cm]{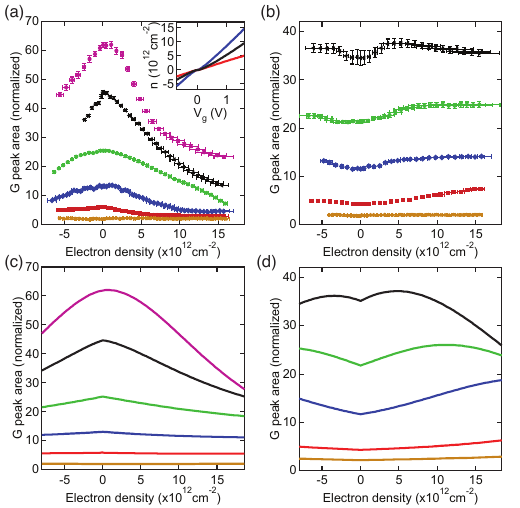} 
\caption{(a) Raman G peak area variation with total charge density for the critical and smaller twist angles.  Purple: 12.6$^{\circ}$, black: 12.1$^{\circ}$, green: 11.6$^{\circ}$, blue: 11.1$^{\circ}$, red: 10.4$^{\circ}$, orange: smaller than 3$^{\circ}$.  Inset is the calculation of bottom (red), top (black) and total (blue) charge densities for the tBLG corresponding to the blue curve in the main figure. (b) Raman G peak area variation with total charge density for twist angles larger than the critical angle.  Black: 13.4$^{\circ}$, green: 13.7$^{\circ}$, blue: 14.1$^{\circ}$, red: 15.1$^{\circ}$, orange: bigger than 16$^{\circ}$.  The uncertainty in the angles is 0.1$^{\circ}$ due to the uncertainty of the G peak area.  The error bars in electron density are from the uncertainty of the geometric capacitance. (c) and (d) Numerical calculation of Raman G peak area variation upon doping for the same twist angles as shown in (a) and (b) respectively.}
\label{fig:results}
\end{center}
\end{figure}

\begin{figure}[h]
\begin{center}
\includegraphics[width=12cm]{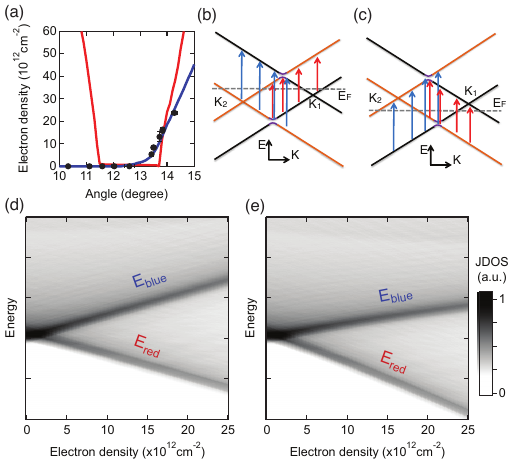} 
\caption{(a) Charge density for the maximum G peak area versus twist angle. The red line is the calculation considering only the interlayer potential imposed by the asymmetric doping in the two layers. The blue line is the calculation including the reduction of Fermi velocity as discussed in the text. (b)-(c) Electronic band structure evolution upon doping of tBLG with electrons and holes respectively. Orange and black Dirac cones are from the top and bottom layers respectively. (d)-(e) JDOS as a function of energy and charge density for (d) only an interlayer potential and (e) an interlayer potential and a reduction of the Fermi velocity.}
\label{fig:TheoreticalCal}
\end{center}
\end{figure}

The Raman G peak area of tBLG with twist angles smaller than 3 degrees or bigger than 16 degrees do not depend on the charge density as shown by the orange markers in Fig.~\ref{fig:results}(a) and Fig.~\ref{fig:results}(b) respectively. So, we focus on the angle range from 10 to 15 degrees where the G peak area depends on charge density. The G peak area of tBLG with a twist angle equal to the critical angle decreases monotonically for both the electron and hole doped regions as shown by the purple markers in Fig.~\ref{fig:results}(a). For all angles smaller than the critical angle, similar behavior is observed as shown by the other markers in Fig.~\ref{fig:results}(a). In contrast, for angles larger than the critical angle, the area initially increases with doping before decreasing at higher doping levels as shown in Fig.~\ref{fig:results}(b). The black curve in Fig.~\ref{fig:results}(b) corresponds to a twist angle of 13.4 $\pm$ 0.1 degrees. Its maximum G peak area occurs at a total charge density of $5.3 \pm 0.9 \times 10^{12}$ cm$^{-2}$. This behavior is observed for either electron or hole doping. The total charge density required to maximize the G peak area for a slightly larger twist angle is much larger. The maximum G peak area occurs at $13 \pm 2 \times 10^{12}$ cm$^{-2}$ for the green curve which corresponds to a twist angle of 13.7 $\pm$ 0.1 degrees. The maxima for the blue and red curves are not yet reached within the charge density range plotted. The charge density when the maximum G peak area occurs as a function of twist angle is shown by the black markers in Fig.~\ref{fig:TheoreticalCal}(a).

%%%%%%%%%%%%%%%
%%%%%%%%%%%%%%%%%%

%The Discussion should be succinct and must not contain subheadings.
To understand this asymmetric angle dependence, we examine the modification of the tBLG band structure upon doping.  In undoped tBLG as shown in the inset of Fig.~\ref{fig:Raman}(a), the Dirac points of the two layers are at the Fermi level since there is no charge doping or interlayer potential.  The two vHs are therefore aligned in momentum exactly between the two Dirac points. However, when the sample is doped, the two Dirac points shift away from the Fermi level asymmetrically\cite{Santos2007,Xian2011,Jose2013,Moon2014,Chiu2014,Chung2015}.  When this occurs, the two vHs are displaced horizontally in momentum as seen in Figs.~\ref{fig:TheoreticalCal}(b) and \ref{fig:TheoreticalCal}(c) for electron and hole doping respectively.  Therefore, as the tBLG is doped the energy separation of the bands splits into two different contributions.  The joint density of states can be defined as JDOS(E) $\propto \int \delta(E_{\mathrm{ab}}(k)-E)\, dk$.  Fig.~\ref{fig:TheoreticalCal}(d) shows the JDOS as a function of energy and charge density considering only the interlayer potential imposed by the asymmetric doping in the two layers. Half of the JDOS becomes lower in energy with increasing doping ($E_{\mathrm{red}}$ branch denoted by red arrows in Fig.~\ref{fig:TheoreticalCal}(b) and ~\ref{fig:TheoreticalCal}(c)) while the other half becomes higher in energy ($E_{\mathrm{blue}}$ branch denoted by blue arrows in Fig.~\ref{fig:TheoreticalCal}(b) and ~\ref{fig:TheoreticalCal}(c)).  Variation of twist angle shifts the $E_{\mathrm{red}}$ and $E_{\mathrm{blue}}$ branches uniformly up or down in energy but does not change their separation.  So, for all angles there should be a charge density when the laser photon energy is resonant with a peak in the JDOS and an enhanced Raman G peak will occur. We use equation~\ref{RamanGpeakCalculation} to calculate the G peak area where the part $\Delta E(k)$ depends on charge density since the band structure is changing. The red curve in Fig.~\ref{fig:TheoreticalCal}(a) is the theoretical calculation for the charge density when the Raman G peak area is a maximum considering only the interlayer potential imposed by the asymmetric doping in the two layers as in Fig.~\ref{fig:TheoreticalCal}(d). Indeed, we find the maximum of the G peak area occurs at finite charge densities for both smaller and larger twist angles due to the resonant scattering pathways discussed above.  Near the critical angle, the maximum G peak area occurs at zero charge density because of the large initial drop in the JDOS with doping which overwhelms the resonant effect.  The charge densities calculated are symmetric about the critical angle and are much bigger than experimentally observed. Therefore, the interlayer potential alone can not explain the asymmetric angle dependence that is observed in the experiment.  To explain the results, there must be some mechanism that breaks the symmetry between the smaller and larger twist angles.

%%%%%%%%%%%%%%%
%%%%%%%%%%%%%%%%%%

To explain the data shown in Figs.~\ref{fig:results} (a) and (b) for different twist angles, we must have a change in the band structure that depends on charge doping. In monolayer graphene, the Fermi velocity reduces upon charge doping due to electron-electron interactions\cite{Basov2008,Geim2011,Elton2013,Yu2013}.  Theoretical work for tBLG has also predicted a reduction\cite{Xian2011} but it has not been experimentally explored.  A reducing Fermi velocity causes the slope of the bands in Fig.~\ref{fig:TheoreticalCal}(b) and \ref{fig:TheoreticalCal}(c) to decrease with doping, which results in a shortening of all scattering pathways. This causes both peaks in the JDOS to move toward lower energy as shown in Fig.~\ref{fig:TheoreticalCal}(e).  It prevents the $E_{\mathrm{blue}}$ branch for smaller twist angles from increasing to match $E_{\mathrm{ex}}$.  Likewise, it causes the $E_{\mathrm{red}}$ branch for bigger twist angles to decrease faster and match $E_{\mathrm{ex}}$ at lower charge densities.  Therefore, the Raman G peak area for twist angles smaller than the critical angle drops monotonically within the charge density range in our experiment since no scattering pathway will be resonant with the excitation photon energy or the effect will be overwhelmed by the large initial drop in the JDOS with doping. On the other hand, the Raman G peak area for bigger twist angles increases to a maximum when $E_{\mathrm{red}}$ matches $E_{\mathrm{ex}}$, overcoming the effect of the large initial drop in the JDOS with doping, then it drops at higher doping levels as $E_{\mathrm{red}}$ becomes smaller than $E_{\mathrm{ex}}$. Here we employ changes in the Fermi velocity as a fitting parameter to explain the qualitative difference between the Raman results as a function of charge density for different twist angles. The blue curve in Fig.~\ref{fig:TheoreticalCal}(a) is the theoretical calculation for the charge density when the maximum G peak area occurs considering both the interlayer potential and a reduction of Fermi velocity. We use a simple linear reduction of the Fermi velocity with total charge doping in the calculation as $v_F = v_{F0} - \alpha n_T $, where $v_{F0}= 0.91 \times 10^6$ m/s is the Fermi velocity of tBLG  at zero charge density which matches the critical twist angle observed experimentally. The slope of the reduction, $\alpha$, is adjusted to match the turning charge densities for the larger twist angles. A reduction of around 2 percent of the Fermi velocity at the largest charge density obtained in the experiment is needed to match our data.  While the exact functional dependence of the Fermi velocity could be something other than a linear decrease with increasing charge density, we have chosen to use this as it is the simplest form and qualitatively explains our data.  In particular, as shown by the red line in Fig.~\ref{fig:TheoreticalCal}(a) a non-changing band structure gives qualitatively different results from our experiments and is symmetric about the critical angle.  A change in the band structure which reduces the energy of the scattering pathways is necessary to explain our experimental results which are asymmetric about the critical angle.

%%%%%%%%%%%%%%%
%\begin{figure}[h]
%\includegraphics[width=8.5cm]{TheoreticalCal.pdf} 
%\caption{Numerical calculation of Raman G peak area variation upon doping for the same twist angles as shown in Fig. 3(a) and 3(b).}
%\label{fig:TheoreticalCal}
%\end{figure}
%%%%%%%%%%%%%%%%%%

To confirm our model for the reduction in Fermi velocity with increasing doping, we calculate the variation of the G peak area with charge density.  This is done using equation~\ref{RamanGpeakCalculation} with a changing separation of the energy levels due to the charge density and a reduction of the Fermi velocity.  Figures~\ref{fig:results}(c) and (d) plot numerical calculations of the Raman G peak area variation with doping for the smaller and larger twist angles corresponding to Figs.~\ref{fig:results}(a) and \ref{fig:results}(b). The calculations agree very well with the experimentally measured curves. They exhibit the distinct difference in density dependence of the smaller and larger twist angles. Furthermore, they demonstrate that the reduction of Fermi velocity plays an important role in the Raman signal of tBLG with doping.  Without the decrease of Fermi velocity with doping, the curves for smaller and larger twist angles would be identical. 
    
In summary, modification of the electronic band structure of tBLG upon doping is realized using a polymer electrolyte top gate. Dirac points from the two layers are shifted apart by an interlayer potential $\Delta\phi$ because of the asymmetric doping causing the two vHs to be misaligned horizontally in momentum. Distinct behavior with charge density is observed for twist angles smaller and larger than the critical angle due to changes in the electronic band structure. From our modeling, a small reduction in the Fermi velocity with doping causes significant changes in the Raman G peak area. We hope that our results will stimulate further theoretical work into the modification of the band structure of twisted bilayer graphene with charge doping.  Our results demonstrate the rich physical properties of tBLG and how they can be modified with doping paving the way for tailoring the electronic band structure of tBLG by external potentials.
\section*{Methods}

%Topical subheadings are allowed. Authors must ensure that their Methods section includes adequate experimental and characterization data necessary for others in the field to reproduce their work.
\subsection*{Raman measurement setup.} Raman spectroscopy was performed on a home built Raman system using a 532 nm Nd:YAG laser under ambient environment and room temperature. The laser was focused onto the sample by a 50$\times$ objective with NA = 0.5, which yielded a spot size smaller than 1.5 $\mu$m. The power of the laser was less than 1 mW in order to avoid damaging the sample. The reflected light passed through a 532 nm high-pass filter before being dispersed by a 600 lines/mm grating.  The resulting spectrum was imaged on a thermoelectrically cooled CCD giving a spectral resolution of about 1 cm$^{-1}$.

\subsection*{Polymer electrolyte preparation.} The polymer electrolyte was composed of lithium perchlorate and polyethylene oxide (PEO) with a weight ratio of 1 to 8 dispersed in methanol and stirred for several hours at 50 $^{\circ}{\rm C}$ until thoroughly dissolved.  A drop of the ion gel was put onto the graphene device and baked for 5 mins at 100 $^{\circ}{\rm C}$ to evaporate the methanol.

\subsection*{Graphene sample synthesis.} The graphene in this study was grown via a low pressure CVD method on  25 $\mu$m thick Cu foil placed within a sealed copper pouch\cite{Li2011}. The pouch was heated to 1040 $^{\circ}{\rm C}$ and annealed for 1 hour under a H$_2$ gas flow rate of 4 sccm at a pressure of 60 mTorr. Then a CH$_4$ flow rate of 1.3 sccm was added to the system for 30 minutes to grow the graphene. The system was quickly cooled down to 350 $^{\circ}{\rm C}$ with the same gas flows as during growth, and then the gas was stopped. 

\bibliography{ReferenceHuang}

%\noindent LaTeX formats citations and references automatically using the bibliography records in your .bib file, which you can edit via the project menu. Use the cite command for an inline citation, e.g.  \cite{Figueredo:2009dg}.

\section*{Acknowledgements}

S.H., M.Y., K.C. and B.J.L. were supported by NSF Career Award No. DMR-0953784 and the U.S. Army Research Laboratory and the U.S. Army Research Office under Contract/Grant Nos. W911NF-09-0333,  W911NF-12-1-0345, and W911NF-14-1-0653. A.S. was supported by NSF award No. PHY-1505556.

\section*{Author contributions statement}

S.H. fabricated the devices and performed the Raman measurements. S.H. and K.C. synthesized the CVD graphene. S.H. and M.Y. performed the STM measurements. A.S. and B.J.L. conceived and provided advice on the experiments. All authors participated in the data discussion and writing of the manuscript. 

\section*{Additional information}
\textbf{Competing financial interests}: The authors declare no competing financial interests.

\end{document}